\documentclass[%
 aip,
 amsmath,amssymb,
 reprint,%
]{revtex4-1}

\usepackage{graphicx}
\usepackage{dcolumn}
\usepackage{bm}

\usepackage[utf8]{inputenc}
\usepackage[T1]{fontenc}
\usepackage{mathptmx}
\usepackage{hyperref}

\usepackage[dvipsnames]{xcolor}

\begin{document}

\preprint{AIP/123-QED}

\title[Hybrid Photonics Lab]{\textcolor{Bittersweet}{Nano-second exciton-polariton lasing in organic microcavities.}
}

\author{A. Putintsev}
\affiliation{ 
Center of Photonics and Quantum Materials, Skolkovo Institute of Science and Technology, Moscow, Russian Federation 121205}
\author{A. Zasedatelev}
\affiliation{ 
Center of Photonics and Quantum Materials, Skolkovo Institute of Science and Technology, Moscow, Russian Federation 121205}
\affiliation{ 
Department of Physics and Astronomy, University of Southampton, Southampton SO17 1BJ, UK}
\author{K. E. McGhee}%
\affiliation{%
Department of Physics and Astronomy, University of Sheffield,
Sheffield S3 7RH, UK}
\author{T. Cookson}%
\affiliation{ 
Department of Physics and Astronomy, University of Southampton, Southampton SO17 1BJ, UK}
\author{K. Georgiou}
\affiliation{%
Department of Physics and Astronomy, University of Sheffield,
Sheffield S3 7RH, UK}
\author{D. Sannikov}
\affiliation{ 
Center of Photonics and Quantum Materials, Skolkovo Institute of Science and Technology, Moscow, Russian Federation 121205}
\author{D. G. Lidzey}
\affiliation{%
Department of Physics and Astronomy, University of Sheffield,
Sheffield S3 7RH, UK}
\author{P. G. Lagoudakis}
\affiliation{ 
Center of Photonics and Quantum Materials, Skolkovo Institute of Science and Technology, Moscow, Russian Federation 121205}
\affiliation{ 
Department of Physics and Astronomy, University of Southampton, Southampton SO17 1BJ, UK}

\date{\today}%

\begin{abstract}
{Organic semiconductors are a promising platform for ambient polaritonics. Several applications, such as polariton routers, and many-body condensed matter phenomena are currently hindered due to the ultra-short polariton lifetimes in organics. Here, we employ a single-shot dispersion imaging technique, using 4 nanosecond long non-resonant excitation pulses, to study polariton lasing in a $\lambda/2$ planar organic microcavity filled with BODIPY-Br dye molecules. At a power threshold density of $1.5 MW/cm^{2}$, we observe the transition to a quasi-steady state, 1.2 ns long-lived, single-mode polariton lasing and the concomitant superlinear increase of photoluminescence, spectral line-narrowing, and energy blueshift.}

\end{abstract}

\maketitle


Strongly coupled organic microcavities are perhaps the most developed material systems for ambient polaritonics. A broad range of suitable materials has enabled the experimental observation of polariton lasing across the whole visible range \cite{KenaCohen2010,Plumhof2014,Daskalakis2014,Dietriche2016,Cookson2017,Scafirimuto2018,Sannikov2019,Rajendran2019,Wei2019,ZasedatelevNature,Yagafarov2020}, as well as device-concepts ranging from ultra-fast transistors and all-optical logic gates \cite{ZasedatelevNature,baranikov2020alloptical}, to  single-photon switching \cite{zasedatelev2020organic}, all at room temperature under ambient conditions. However, the localised nature of Frenkel excitons, alongside the typically ultra-short polariton lifetimes, $\sim100 fs$, limit both the study of many-body phenomena and applications that require macroscopic polariton transport. Unlike the case of inorganic semiconductor microcavities, where continuous wave (CW) excitation allows for the replenishment of particle losses, leading to the realisation of steady-state polariton condensates, in organic semiconductors polaron formation and photobleaching hinders CW operation. In a recent study, quasi-steady operation was implemented using nanosecond non-resonant optical excitation in multi-$\lambda$ fluorescent protein microcavities \cite{Dietriche2016} that led to the observation of extended temporal coherence, reaching $150 ps$ \cite{Betzold2019}, i.e. 3 orders of magnitude longer than the corresponding cavity lifetime. Multi-$\lambda$ cavities host several consecutive cavity modes complicating the distinction between strong and weak coupling regimes. Here, we use a single-mode $\lambda/2$ strongly coupled microcavity of a BODIPY dye molecule. BODIPY dyes have been the subject of extensive studies for their applications in the strong coupling regime \cite{Musser2017}. Strongly coupled BODIPY microcavities \cite{GrantAOM2016,GeorgiouACSPhot2016,LiuB2019} and polariton lasing in these structures allow for highly monochromatic tuneable coherent emission of duration up to $\simeq$ 2 picoseconds \cite{Cookson2017,Sannikov2019,Yagafarov2020}. 

\begin{figure}[h!]
\includegraphics[width=1\linewidth]{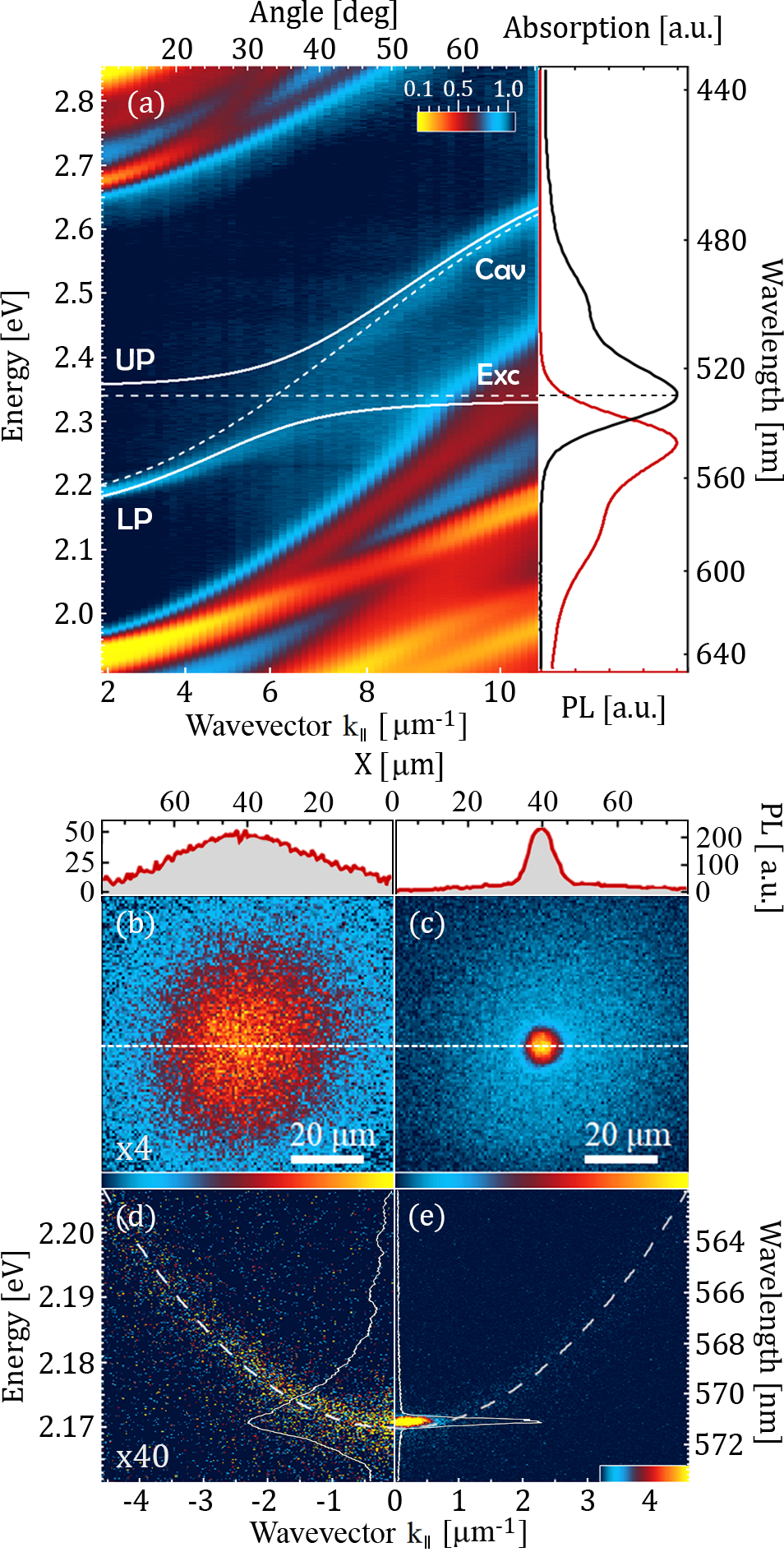}
\vspace{-0.2cm}
\caption{\label{fig:signature} (a) Angularly resolved reflectivity map of the microcavity showing the anticrossing of the bare exciton and cavity modes (Exc, Cav, white dashed-curves) at $k_{||}=5\mu m^{-1}$. The white solid-lines are the calculated upper (UP) and lower (LP) polariton branches for a vacuum Rabi-splitting of $110meV$ and -164 meV exciton-photon detuning. The right panel shows the absorption (black) and photoluminescence (red) of the neat BODIPY-Br film. (b), (c) Single-shot real-space photoluminescence images showing the spatial distribution of polariton density under a single non-resonant nanosecond excitation pulse below and above condensation threshold, respectively. The top panels show the respective cross-section of the photoluminescence at the position of the dashed white lines. (d), (e) Single-shot dispersion imaging of the \textit{$E,k_{||}$}-distributions of polariton density below and above the condensation threshold respectively. The white dashed-lines show the position of the lower polariton branch and the solid-lines the angularly integrated photoluminescence. x4 and x40 are the scale-factors for normalized colorbars.}
\vspace{-0.7cm}
\end{figure}

In this letter, we demonstrate quasi-steady state polariton lasing in a single-mode $\lambda/2$ strongly coupled BODIPY microcavity, namely the nanosecond long-lasting polariton condensate which exceeds the cavity lifetime ($\sim100fs$) by 4-orders of magnitude. We employ a single-shot dispersion, real-space imaging, and time-resolved photoluminescence (PL) technique that allows for full characterisation of the polariton condensate emission and its transient PL dynamics under a single excitation pulse, where each excitation pulse-amplitude and spatial profile are simultaneously recorded. The aforementioned technique prevents averaging of the polariton emission characteristics due to pulse-to-pulse intensity fluctuations ($\sim 15\%$ for nanosecond YAG pumped optical parametric amplifiers) and photo-bleaching of the organic emitter. The microcavity structure under investigation consists of a $\lambda/2$-thin organic slab based on BODIPY-Br dye molecules dispersed into polystyrene host as 1:10 by mass that, in turn, is sandwiched between bottom and top distributed Bragg reflectors of 10 and 8 alternating pairs of Nb$_{2}$O$_{5}-$ and SiO$_{2}-$ $\lambda/4$ layers, respectively. The microcavity has a Q-factor of $\sim$450, supporting strong light-matter coupling between the cavity photon and the Frenkel exciton resonance associated with singlet $S_{0,0}-S_{1,0}$ electronic transition centered at 2.34 eV. Figure \ref{fig:signature}a shows the anglularly resolved reflectivity spectrum that clearly demonstrates strong coupling at room temperature as evidenced by the anti-crossing of the bare exciton and cavity modes at $k_{||}=5\mu m^{-1}$ resulting in a vacuum Rabi-splitting of $110meV$. We fit the reflectivity map with modes from coupled oscillator model where bare cavity and excitonic resonances and new hybridized states of lower and upper polariton states are shown with white dashed and solid lines, respectively. The energy of excitonic resonance used in the model corresponds to the peak absorption of the active material with its absorption and PL spectra shown on the right inset of Figure \ref{fig:signature}a. 

We optically excite the microcavity using $4 ns$ single pulses tuned at the first Bragg minimum of the reflectivity spectrum ($462nm$) in order to reduce sample heating. We record single-shot real-space images and \textit{$E,k_{||}$}-distributions of polariton PL in transmission configuration. Figure \ref{fig:signature}b shows a typical linear polariton PL real-space image at $P = 0.75 MW/cm^{2}$; the cross-section along \textit{x}-axis has the same full width at half maximum (FWHM) with the pump beam ($40\mu m$). With increased pump power by a factor of $\sim$2 ($P = 1.7 MW/cm^{2}$) the spatial distribution shrinks substantially ($8\mu m$ in FWHM), as shown in Fig.\ref{fig:signature}c, and the emission intensity increases nonlinearly. Figures \ref{fig:signature}d,e are single-shot dispersion images showing the polariton PL distribution along the lower polariton branch (LPB), for the two pumping intensities. At low pump power, polariton PL is distributed along the LPB, while for $\sim$ twice the pump power we observe a collapse of the polariton distribution at the ground polariton state. The white line profiles show the angularly integrated PL evidencing strong linewidth narrowing and an energy blueshift at the higher pump power. These spectroscopic characteristics are indicative of polariton condensation in semiconductor microcavities \cite{Plumhof2014,Daskalakis2014,Dietriche2016,Cookson2017,Scafirimuto2018,Wei2019,Sannikov2019,Rajendran2019,ZasedatelevNature,Yagafarov2020}. Next we perform a full pump fluence dependence to resolve the threshold excitation density for polariton condensation.   
\begin{figure}[t!]
\includegraphics[width=1\linewidth]{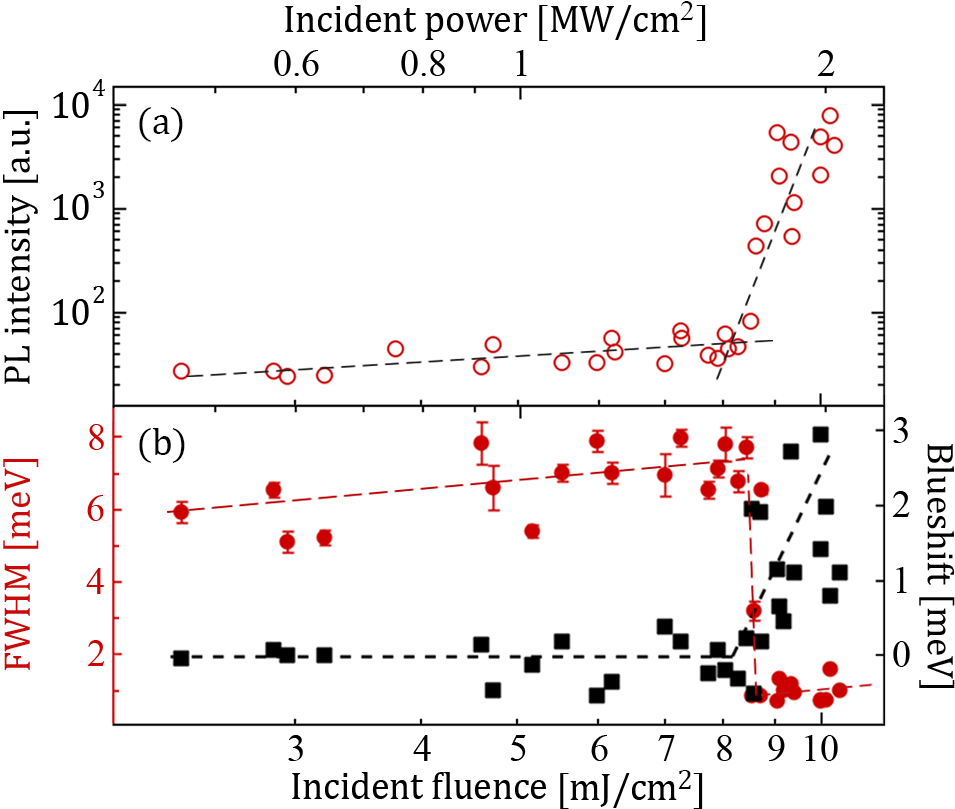}
\setlength{\belowcaptionskip}{-10pt}
\vspace{-0.5cm}
\caption{\label{fig:blueshift} (a) Polariton photoluminescence intensity at $k_{\|}\sim 0$ integrated into the range over $\pm0.2\mu m^{-1}$  as the function of incident pump fluence (bottom axis), incident pump power density (top axis). The crossing point of the black dashed lines is used to define condensation threshold. (b) The linewidth of polariton photoluminescence at FWHM (solid red circles) and the energy shift of the lower polariton mode at $k_{\|}=0$ (solid black squares) versus incident pump fluence (bottom axis) and incident pump power density (top axis). The dashed lines are guides to the eye.}
\vspace{-0.2cm}
\end{figure}

\begin{figure*}
\includegraphics[width=1\linewidth]{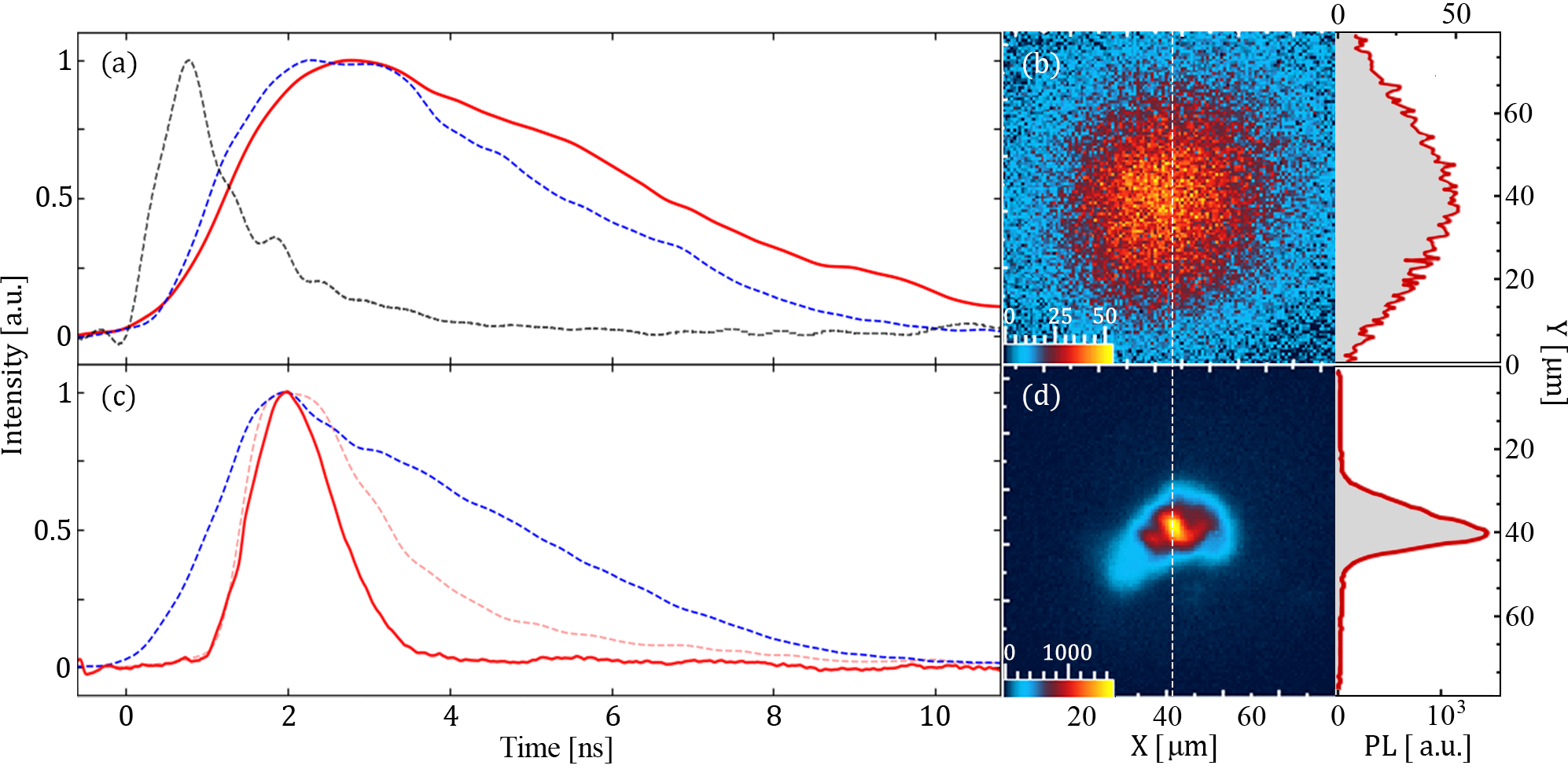}
\vspace{-0.5cm}
\caption{\label{fig:TimeResolved} (a) Normalized, time-resolved polariton photoluminescence intensity (red solid-line)  below condensation threshold, averaged over five realisations/pulses. The blue dashed-line shows the temporal profile of a single excitation pump pulse, and the black dashed-line corresponds to the instrument response function (IRF) of the detector. (b) Single-shot real-space polariton photoluminescence image below condensation threshold. (c) Single-shot, normalized, time-resolved polariton photoluminescence intensity (red dashed-line) at approximately twice the condensation threshold and its deconvolution with IRF (red solid-line). The blue dashed-line corresponds to the normalized single-shot pump pulse profile. (d) Single-shot real-space polariton photoluminescence image at approximately twice the condensation threshold. The right panels of (b,d) show the profile of the photoluminescence intensity at the position of the vertical white dashed-line.}
\vspace{-0.3cm}
\end{figure*}

We perform single-shot dispersion imaging for a range of pump fluences. Figure \ref{fig:blueshift}a,b show  the pump-power dependencies of PL intensity, linewidth, and blueshift at the ground polariton state $k_{||}=0$. We observe a clear threshold of the emission intensity at an incident pump fluence of $8 mJ/cm^{2}$ which is almost coincides with the threshold ($6 mJ/cm^{2}$) under 2ps-short pulsed excitation\cite{Sannikov2019}. The observed condensation threshold is equivalent to $1.5 MW/cm^{2}$ in terms of irradiance - the conventional parameter for the quasi--CW excitation; comparing thresholds in inorganic and hybrid organic-inorganic systems operating under qusi--CW pumping at room temperature\cite{Lu12,Paschos2017}, it requires $\sim50-times$ higher pump power. We have defined threshold as the intersection point of linear and nonlinear curves fitting the PL intensity, shown in Fig. \ref{fig:blueshift}a by dashed-lines. At threshold pump fluence we observe a 7-fold reduction in linewidth from 7 meV to 1 meV and a concomitant blueshift of the ground polariton state up to 3 meV, as shown in Fig. \ref{fig:blueshift}b red and black data, respectively. We believe the condensate linewidth of 1 meV is the result of inhomogeneous broadening due to dynamic blueshift. Recently, we have shown the blueshift arises from the interplay of the saturation effects and intermolecular energy migration of both: bight and dark molecules\cite{Yagafarov2020}. These spectroscopic characteristics together with the spatial collapse of the polariton emission provide strong experimental evidence for the formation of a polariton condensate under nanosecond optical excitation. Interestingly, we find that the spectral properties of polariton condensates under nanosecond excitation are in quantitative agreement with previous studies using picosecond excitation pulses \cite{Cookson2017,Sannikov2019}. It remains therefore to resolve the temporal dynamics of the emission and identify whether the duration of the condensate extends substantially beyond the cavity lifetime.    


We perform single-shot time-resolved PL measurements using a silicon photodiode with $0.5 ns$ rise-time. Figure \ref{fig:TimeResolved}a shows the normalized intensity of the temporal profile of a typical pump pulse (blue dashed-line), the instrument response function (IRF) (black dotted-line), and polariton PL (red solid-line) averaged over five pulses to improve signal-to-noise ratio. For convenience, we define $0 ns$ time at the beginning of IRF's rising edge. The IRF is limited to $\simeq$ $1.03 ns$ at FWHM and the excitation pulse has a duration of $\simeq 4 ns$ at FWHM. The polariton PL is collected within the whole numerical aperture of the microscope objective ($NA=0.42$). Below threshold, at $P\simeq 0.5\times P_{th}$ the temporal profile of polariton PL follows the pump pulse, extending to $\simeq 5.5 ns$ at FWHM. The dominant process populating polariton states is intracavity radiative pumping as follows from the extensive study of polariton population mechanisms in such dye-filled microcavities\cite{GrantAOM2016}. It implies an effective population of polariton states through the radiative decay of fluorescent dyes inside the cavity.
Therefore, the lifetime of the exciton reservoir governs relaxation dynamics of polariton PL below the threshold and defines time delay of about 1ns with respect to the pump profile accordingly. It agrees well with the typical PL decay time of bare BODIPY-Br films \cite{Musser2017}. Since the energy difference between exciton reservoir and the ground polariton state is equal to 164meV one can also anticipate contribution from vibronic relaxation channel\cite{Tartakovskii2001,AdFM2011,Somachi2011,ZasedatelevNature}. Figure \ref{fig:TimeResolved}b shows a single-shot spatial profile of polariton PL for one of the pulses included in the average of the temporal polariton PL profile of Figure \ref{fig:TimeResolved}a. Figure \ref{fig:TimeResolved}c shows the original and deconvoluted temporal profiles of the normalised polariton PL for pump fluence twice above the threshold ($P\simeq 2\times P_{th}$) with respect to the pump pulse. We observe a shortening of the rise time of polariton PL above threshold, resolution limited by the IRF, revealing the rapid formation of the condensate due to bosonic stimulation of carriers from the exciton reservoir to the ground polariton state. The condensate lifetime defined as the duration of deconvoluted polariton PL at FWHM is equal to $\simeq 1.2 ns$, that exceeds cavity lifetime by a factor of $\sim 10000$. Figure \ref{fig:TimeResolved}d shows the corresponding single-shot spatial condensate profile, which has collapsed below the spatial extend of the excitation pulse. The collapse originates from the nonlinear increase of PL at the area within the pump profile, where the optical gain exceeds losses. Asymmetries in the condensate profile presumably relate to the effect of energy disorder inherent to organic microcavities. The small spatial distribution of the condensate reduces dynamical instabilities in agreement with previous studies\cite{KenaCohen2015,Bobrovskaya}. Comparison of the spatial PL profiles of Figures \ref{fig:TimeResolved}b,d provide supporting evidence that the temporal profile of Fig.\ref{fig:TimeResolved}c corresponds to the emission from a polariton condensate. To the best of our knowledge, the temporal dynamics of polariton PL above condensation threshold provide the first direct observation of a quasi-steady state polariton condensate in a $\lambda/2$ planar organic microcavity. 

The direct measurement of the condensate lifetime provides valuable insight on the transient processes of nonequilibrium polariton condensation, its connection with spectral and polarization properties \cite{Yagafarov2020}, dynamical instabilities, and noise limits of coherent polariton sources \cite{Bobrovskaya}. The latter study elucidates the crucial role of a condensate lifetime with respect to the Bogoliubov instability time-scale on the onset of domain formation in a condensate wavefunction that dictates whether it posses homogeneous or highly disordered spatial distribution. Moreover, characterization of the condensate lifetime facilitates the understanding of complex nonlinear dynamics, including rates of depletion and replenishment of the exciton reservoir and polariton states, respectively, under pulsed excitation \cite{Yagafarov2020}. Long-lasting condensates exceeding polartion lifetime for several orders of magnitude push the system one step closer towards the regime of dynamic equilibrium as it corresponds to quasi steady-state operation regime in view of ultrafast dynamics inherent to strongly-coupled organic microcavities. This regime allows one to investigate hydrodynamics of polaritons \cite{RevModPhys.85.299}, including room-temperature superfluidity \cite{Lerario2017} and interactions between separate condensates \cite{dusel2019roomWirzburg} that are capable of flowing over sufficient distances within the condensate lifetime. Realisation of nanosecond organic polariton condensates offers the possibility for devices relying on connected polariton condensates, paving the way for all-optical polariton circuitry on a chip \cite{Gao2012,Nguyen2013,Ballarini2013,Sturm2014,Marsault2015}, available now in ambient conditions \cite{ZasedatelevNature,zasedatelev2020organic,baranikov2020alloptical}.

\begin{acknowledgments}
We acknowledge the support of the UK’s Engineering and Physical Sciences Research Council (grant EP/M025330/1 on Hybrid Polaritonics), the RFBR projects No. 20-52-12026 (jointly with DFG) and No. 20-02-00919. A.P. and A.Z. acknowledge financial support from the Russian Scientific Foundation (RSF) grant.
\end{acknowledgments}

\section*{Data availability}
The data that support the findings of this study are available from the University of Southampton repository at https://doi.org/10.5258/SOTON/D1374.

\nocite{*}
\bibliography{ArXiv}
\end{document}